\documentclass[12pt,a4paper]{article}

\usepackage{ifthen} 
\newboolean{pdflatex}
\setboolean{pdflatex}{true} 

\newboolean{articletitles}
\setboolean{articletitles}{true} 

\newboolean{uprightparticles}
\setboolean{uprightparticles}{false} 

\newboolean{inbibliography}
\setboolean{inbibliography}{false} 

\usepackage{microtype}
\usepackage{lineno}  
\usepackage{xspace} 
\usepackage{caption} 

\usepackage{graphicx}  
\usepackage{color}
\usepackage{colortbl}
\graphicspath{{./figs/}} 

\usepackage{amsmath} 
\usepackage{amssymb}
\usepackage{amsfonts}
\usepackage{upgreek} 

\newcommand*\patchAmsMathEnvironmentForLineno[1]{%
\expandafter\let\csname old#1\expandafter\endcsname\csname #1\endcsname
\expandafter\let\csname oldend#1\expandafter\endcsname\csname
end#1\endcsname
 \renewenvironment{#1}%
   {\linenomath\csname old#1\endcsname}%
   {\csname oldend#1\endcsname\endlinenomath}%
}
\newcommand*\patchBothAmsMathEnvironmentsForLineno[1]{%
  \patchAmsMathEnvironmentForLineno{#1}%
  \patchAmsMathEnvironmentForLineno{#1*}%
}
\AtBeginDocument{%
\patchBothAmsMathEnvironmentsForLineno{equation}%
\patchBothAmsMathEnvironmentsForLineno{align}%
\patchBothAmsMathEnvironmentsForLineno{flalign}%
\patchBothAmsMathEnvironmentsForLineno{alignat}%
\patchBothAmsMathEnvironmentsForLineno{gather}%
\patchBothAmsMathEnvironmentsForLineno{multline}%
\patchBothAmsMathEnvironmentsForLineno{eqnarray}%
}

\usepackage{hyperref}    
\usepackage[all]{hypcap} 

\def\MagUp {\mbox{\em Mag\kern -0.05em Up}\xspace}

\ifthenelse{\boolean{uprightparticles}}%
{

 \def\Pmu         {\ensuremath{\upmu}\xspace}

 \def\Ppi         {\ensuremath{\uppi}\xspace}

 \def\Ptau        {\ensuremath{\uptau}\xspace}

 \def\Ppsi        {\ensuremath{\uppsi}\xspace}

 \def\PDelta      {\ensuremath{\Delta}\xspace}                 
 \def\PXi      {\ensuremath{\Xi}\xspace}                 
 \def\PLambda      {\ensuremath{\Lambda}\xspace}                 
 \def\PSigma      {\ensuremath{\Sigma}\xspace}                 
 \def\POmega      {\ensuremath{\Omega}\xspace}                 
 \def\PUpsilon      {\ensuremath{\Upsilon}\xspace}                 
 
 \def\PB      {\ensuremath{\mathrm{B}}\xspace}                 
                  
 \def\PD      {\ensuremath{\mathrm{D}}\xspace}

 \def\PJ      {\ensuremath{\mathrm{J}}\xspace}                 
 \def\PK      {\ensuremath{\mathrm{K}}\xspace}

 \def\Pb      {\ensuremath{\mathrm{b}}\xspace}                 
 \def\Pc      {\ensuremath{\mathrm{c}}\xspace}                 
                  
 \def\Pe      {\ensuremath{\mathrm{e}}\xspace}

 \def\Pi      {\ensuremath{\mathrm{i}}\xspace}

 \def\Pp      {\ensuremath{\mathrm{p}}\xspace}

 \def\Ps      {\ensuremath{\mathrm{s}}\xspace}

}
{

 \def\Pmu         {\ensuremath{\mu}\xspace}

 \def\Ppi         {\ensuremath{\pi}\xspace}

 \def\Ptau        {\ensuremath{\tau}\xspace}

 \def\Ppsi        {\ensuremath{\psi}\xspace}                 
                  
 \mathchardef\PDelta="7101
 \mathchardef\PXi="7104
 \mathchardef\PLambda="7103
 \mathchardef\PSigma="7106
 \mathchardef\POmega="710A
 \mathchardef\PUpsilon="7107
                  
 \def\PB      {\ensuremath{B}\xspace}                 
                  
 \def\PD      {\ensuremath{D}\xspace}

 \def\PJ      {\ensuremath{J}\xspace}                 
 \def\PK      {\ensuremath{K}\xspace}

 \def\Pb      {\ensuremath{b}\xspace}                 
 \def\Pc      {\ensuremath{c}\xspace}                 
                  
 \def\Pe      {\ensuremath{e}\xspace}

 \def\Pi      {\ensuremath{i}\xspace}

 \def\Pp      {\ensuremath{p}\xspace}

 \def\Ps      {\ensuremath{s}\xspace}

}

\makeatletter
\ifcase \@ptsize \relax
  \newcommand{\miniscule}{\@setfontsize\miniscule{4}{5}}
\or
  \newcommand{\miniscule}{\@setfontsize\miniscule{5}{6}}
\or
  \newcommand{\miniscule}{\@setfontsize\miniscule{5}{6}}
\fi
\makeatother

\DeclareRobustCommand{\optbar}[1]{\shortstack{{\miniscule (\rule[.5ex]{1.25em}{.18mm})}
  \\ [-.7ex] $#1$}}

\def\epem       {{\ensuremath{\Pe^+\Pe^-}}\xspace}

\def\mup        {{\ensuremath{\Pmu^+}}\xspace}
\def\mun        {{\ensuremath{\Pmu^-}}\xspace} 
\def\mumu       {{\ensuremath{\Pmu^+\Pmu^-}}\xspace}

\def\tautau     {{\ensuremath{\Ptau^+\Ptau^-}}\xspace}

\def\squark    {{\ensuremath{\Ps}}\xspace}

\def\cquark    {{\ensuremath{\Pc}}\xspace}

\def\bquark    {{\ensuremath{\Pb}}\xspace}
\def\bquarkbar {{\ensuremath{\overline \bquark}}\xspace}
\def\bbbar     {{\ensuremath{\bquark\bquarkbar}}\xspace}

\def\pion   {{\ensuremath{\Ppi}}\xspace}
\def\piz    {{\ensuremath{\pion^0}}\xspace}

\def\pip    {{\ensuremath{\pion^+}}\xspace}
\def\pim    {{\ensuremath{\pion^-}}\xspace}

\def\kaon    {{\ensuremath{\PK}}\xspace}
  \def\Kbar    {{\kern 0.2em\overline{\kern -0.2em \PK}{}}\xspace}

\def\KorKbar    {\kern 0.18em\optbar{\kern -0.18em K}{}\xspace}

\def\Kp      {{\ensuremath{\kaon^+}}\xspace}
\def\Km      {{\ensuremath{\kaon^-}}\xspace}

  \def\Dbar    {{\kern 0.2em\overline{\kern -0.2em \PD}{}}\xspace}
\def\D       {{\ensuremath{\PD}}\xspace}

\def\DorDbar    {\kern 0.18em\optbar{\kern -0.18em D}{}\xspace}
\def\DtwoorDtwobar {\kern -0.25em\optbar{\kern 0.25em D_2^*}{}\xspace}
\def\Dz      {{\ensuremath{\D^0}}\xspace}

\def\Dp      {{\ensuremath{\D^+}}\xspace}

\def\Dstar   {{\ensuremath{\D^*}}\xspace}

\def\Dstarz  {{\ensuremath{\D^{*0}}}\xspace}
\def\DzorDstarz  {{\ensuremath{\D^{(*)0}}}\xspace}
\def\Dstarzb {{\ensuremath{\Dbar{}^{*0}}}\xspace}
\def\Dstarp  {{\ensuremath{\D^{*+}}}\xspace}

\def\Dssp    {{\ensuremath{\D^{*+}_\squark}}\xspace}

\def\B       {{\ensuremath{\PB}}\xspace}

\def\Bbar    {{\ensuremath{\kern 0.18em\overline{\kern -0.18em \PB}{}}}\xspace}

\def\BorBbar    {\kern 0.18em\optbar{\kern -0.18em B}{}\xspace}
\def\Bz      {{\ensuremath{\B^0}}\xspace}
\def\Bstarz      {{\ensuremath{\B^{*0}}}\xspace}
\def\Bstarp      {{\ensuremath{\B^{*+}}}\xspace}
\def\BzorBstarz      {{\ensuremath{\B^{(*)0}}}\xspace}

\def\BzorBzbar  {\kern 0.18em\optbar{\kern -0.18em B}{}^0\xspace}
\def\Bu      {{\ensuremath{\B^+}}\xspace}
\def\Bub     {{\ensuremath{\B^-}}\xspace}
\def\Bp      {{\ensuremath{\Bu}}\xspace}
\def\Bm      {{\ensuremath{\Bub}}\xspace}

\def\Bd      {{\ensuremath{\B^0}}\xspace}
\def\Bs      {{\ensuremath{\B^0_\squark}}\xspace}
\def\Bds      {{\ensuremath{\B^0_{(\squark)}}}\xspace}
\def\Bsstar  {{\ensuremath{\B^{*0}_\squark}}\xspace} 
\def\BsorBsstar  {{\ensuremath{\B^{(*)0}_\squark}}\xspace} 
\def\BorBs  {{\ensuremath{\B^{0}_{(\squark)}}}\xspace} 
\def\BorBsstar  {{\ensuremath{\B^{*0}_{(\squark)}}}\xspace} 

\def\Bc      {{\ensuremath{\B_\cquark^+}}\xspace}
\def\Bcp     {{\ensuremath{\B_\cquark^+}}\xspace}

\def\jpsi     {{\ensuremath{{\PJ\mskip -3mu/\mskip -2mu\Ppsi\mskip 2mu}}}\xspace}

  \def\Y#1S{\ensuremath{\PUpsilon{(#1S)}}\xspace}

\def\proton      {{\ensuremath{\Pp}}\xspace}

\def\Lz          {{\ensuremath{\PLambda}}\xspace}
\def\Lbar        {{\ensuremath{\kern 0.1em\overline{\kern -0.1em\PLambda}}}\xspace}
\def\LorLbar    {\kern 0.18em\optbar{\kern -0.18em \PLambda}{}\xspace}

\def\Lb      {{\ensuremath{\Lz^0_\bquark}}\xspace}

\def\BF         {{\ensuremath{\cal B}}\xspace}

\newcommand{\decay}[2]{\ensuremath{#1\!\to #2}\xspace}         

\def\to                 {\ensuremath{\rightarrow}\xspace}

\def\qsq       {{\ensuremath{q^2}}\xspace}

\def\AT#1     {\ensuremath{A_{\mathrm{T}}^{#1}}\xspace}           

\def\C#1      {\ensuremath{\mathcal{C}_{#1}}\xspace}                       
\def\Cp#1     {\ensuremath{\mathcal{C}_{#1}^{'}}\xspace}                    
\def\Ceff#1   {\ensuremath{\mathcal{C}_{#1}^{\mathrm{(eff)}}}\xspace}        
\def\Cpeff#1  {\ensuremath{\mathcal{C}_{#1}^{'\mathrm{(eff)}}}\xspace}       
\def\Ope#1    {\ensuremath{\mathcal{O}_{#1}}\xspace}                       
\def\Opep#1   {\ensuremath{\mathcal{O}_{#1}^{'}}\xspace}                    

\newcommand{\tev}{\ifthenelse{\boolean{inbibliography}}{\ensuremath{~T\kern -0.05em eV}\xspace}{\ensuremath{\mathrm{\,Te\kern -0.1em V}}}\xspace}
\newcommand{\gev}{\ensuremath{\mathrm{\,Ge\kern -0.1em V}}\xspace}
\newcommand{\mev}{\ensuremath{\mathrm{\,Me\kern -0.1em V}}\xspace}
\newcommand{\kev}{\ensuremath{\mathrm{\,ke\kern -0.1em V}}\xspace}
\newcommand{\ev}{\ensuremath{\mathrm{\,e\kern -0.1em V}}\xspace}
\newcommand{\gevc}{\ensuremath{{\mathrm{\,Ge\kern -0.1em V\!/}c}}\xspace}
\newcommand{\mevc}{\ensuremath{{\mathrm{\,Me\kern -0.1em V\!/}c}}\xspace}
\newcommand{\gevcc}{\ensuremath{{\mathrm{\,Ge\kern -0.1em V\!/}c^2}}\xspace}
\newcommand{\gevgevcccc}{\ensuremath{{\mathrm{\,Ge\kern -0.1em V^2\!/}c^4}}\xspace}
\newcommand{\mevcc}{\ensuremath{{\mathrm{\,Me\kern -0.1em V\!/}c^2}}\xspace}

\def\mub{\ensuremath{{\rm \,\upmu b}}\xspace}

\def\invfb   {\ensuremath{\mbox{\,fb}^{-1}}\xspace}

\def\ps   {\ensuremath{{\rm \,ps}}\xspace}

\def\gsim{{~\raise.15em\hbox{$>$}\kern-.85em
          \lower.35em\hbox{$\sim$}~}\xspace}
\def\lsim{{~\raise.15em\hbox{$<$}\kern-.85em
          \lower.35em\hbox{$\sim$}~}\xspace}

\def\root       {\mbox{\textsc{Root}}\xspace}

\def\tell1  {TELL1\xspace}
\def\ukl1   {UKL1\xspace}

\newcommand{\eg}{\mbox{\itshape e.g.}\xspace}
\newcommand{\ie}{\mbox{\itshape i.e.}\xspace}

\usepackage{cite} 
\usepackage{mciteplus}

\usepackage{rotating} 
\usepackage{arydshln} 
\usepackage{relsize} 
\usepackage[hang,flushmargin]{footmisc}

\begin{document}

\renewcommand{\thefootnote}{\fnsymbol{footnote}}
\setcounter{footnote}{1}

\begin{titlepage}
\pagenumbering{roman}

{\bf\boldmath\huge
\begin{center}
  Prospects for studies of $\Dstar^0 \to \mumu$ and $\BorBsstar \to \mumu$ decays
\end{center}
}

\vspace*{1.5cm}

\begin{center}
F.~Abudin{\'e}n$^1$,  T.~Blake$^1$, U.~Egede$^2$, T.~Gershon$^1$
\bigskip\\
{\it\footnotesize 
$ ^1$ Department of Physics, University of Warwick, Coventry, United Kingdom\\
$ ^2$ School of Physics and Astronomy, Monash University, Melbourne, Australia\\
}
\end{center}

\vspace{\fill}

\begin{abstract}
  \noindent
  Weak decays of the vector $\Dstar^0$ and $\BorBsstar$ mesons to the $\mumu$ final state provide novel potential to test the Standard Model of particle physics.
  Such processes have extremely small branching fractions as the vector mesons are able to decay through electromagnetic and (for the  $\Dstar^0$ meson) strong interactions.  
  Nonetheless, the production of copious quantities of these particles in LHC collisions, and the ability to exploit experimental techniques that can suppress background to low levels, provides good potential to reach interesting sensitivity.
  The possibility to reconstruct these processes as part of the decay chain of \Bm\ or \Bc\ mesons appears particularly attractive due to the clean experimental signature of the displaced vertex.
  Indeed, published LHCb data on $\Bm \to \pim\mumu$ decays already implies a stringent limit on the branching fraction of $\Dstarz \to \mumu$.
  Estimates are made on the achievable sensitivity to $\Dstarz \to \mumu$ and $\BorBsstar \to \mumu$ decays with the LHCb experiment.
\end{abstract}

\vspace{\fill}

\end{titlepage}

\newpage
\setcounter{page}{2}
\mbox{~}

\clearpage

\renewcommand{\thefootnote}{\arabic{footnote}}
\setcounter{footnote}{0}

\pagestyle{plain} 
\setcounter{page}{1}
\pagenumbering{arabic}

\section*{Introduction}
Weak decays of heavy-flavoured hadrons provide a range of methods to test the Standard Model (SM) of particle physics.
In particular, many such transitions are suppressed by particular features of the SM such as the GIM mechanism~\cite{Glashow:1970gm} and the CKM quark mixing matrix~\cite{Cabibbo:1963yz,Kobayashi:1973fv}.
As a consequence, the SM predicts a distinctive pattern of decay rates to various different final states, which may be modified by contributions from physics beyond the SM.
Experimental and theoretical investigations in this area are therefore a priority in the field.

Until now, experimental studies of weak decays have been almost completely limited to the ground-state hadrons; considering neutral heavy-flavoured mesons, these are the pseudoscalar $\Dz$, $\Bz$ and $\Bs$ states.
The leptonic decays, to the $\ell^+\ell^-$ final state where $\ell = e, \mu$ and (for \BorBs\ decays) $\tau$, have branching fractions that are suppressed by the chiral structure of the SM weak interaction, and that can be predicted with small theoretical uncertainties~\cite{Bobeth:2013uxa}.
These features together make them highly sensitive to potential contributions from physics beyond the SM.  
Intense activity on the $\BorBs \to \mumu$ channels has resulted in the observation of the $\Bs \to \mumu$ decay by the LHCb, CMS and ATLAS experiments, and a combined limit on the $\Bd \to \mumu$ branching fraction that approaches its SM value~\cite{LHCb-PAPER-2017-001,Sirunyan:2019xdu,Aaboud:2018mst,LHCb-PAPER-2021-007,LHCb-PAPER-2021-008,LHCb-CONF-2020-002}.
The experimental limits on ${\cal B}\left(\Dz \to \epem\right)$~\cite{Belle:2010ouj}, ${\cal B}\left(\Dz \to \mumu\right)$~\cite{LHCb-PAPER-2013-013}, ${\cal B}\left(\Bds \to \epem\right)$~\cite{LHCb-PAPER-2020-001} and ${\cal B}\left(\Bds \to \tautau\right)$~\cite{LHCb-PAPER-2017-003} are still several orders of magnitude above their SM predictions.

It is also possible to consider weak decays of the excited counterparts of the pseudoscalar mesons, the vector $\Dstarz$, $\Bstarz$ and $\Bsstar$ resonances.
In contrast to the situation for pseudoscalar mesons, the leptonic vector meson decays have no chiral suppression.
Consequently the decay widths for each of the $\ell^+\ell^-$ final states are expected to be the same, in the SM, up to effects related to the lepton mass (\eg, the available phase space), and will be larger compared to those for the pseudoscalar decays.
However, the vector mesons can also decay via electromagnetic and (for the $\Dstarz$ meson) strong transitions, which have widths many orders of magnitude larger than those for the weak decays.
Therefore, the branching fractions of the weak decays are suppressed to what would usually be considered unobservably small levels.
As an illustration, one can compare the width of the $\Dstarp$ vector state, $\Gamma_{\Dstarp} = 83 \kev$~\cite{Lees:2013zna}, with that of its $\Dp$ pseudoscalar counterpart, $\Gamma_{\Dp} = \hbar/\tau_{\Dp} \approx \hbar/(1.0 \ps) = 0.7~{\rm meV}$~\cite{Link:2002bx}, a difference of over 8 orders of magnitude.\footnote{
  A similar exercise for the $\DzorDstarz$, $\BzorBstarz$ and $\BsorBsstar$ states is not possible because the widths of the vector resonances have not been measured and have significant theoretical uncertainty.
}
This is in the ballpark that one would naively expect, given that the weak decays are suppressed by the Fermi constant, but have more phase space available compared to the electromagnetic and strong transitions in which flavour is conserved.
Given that weak decays lead to a plethora of different final states, the branching fractions for even the most favoured are unlikely to be above $10^{-9}$, unless large enhancement factors due to physics beyond the SM are at play.

Hints of physics beyond the SM in $B$ meson decays have, nevertheless, prompted theoretical activity on weak decays of excited heavy-flavoured states.
Collectively referred to as the flavour anomalies, these hints include tensions between SM predictions and experimental measurements in branching fractions and angular observables in decays mediated by $b \to s \ell^+\ell^-$ transitions, including observables that are sensitive to violations of lepton universality (see Ref.~\cite{Albrecht:2021tul} for a recent review).
The leptonic $\Dstarz$, $\Bstarz$ and $\Bsstar$ decays are of most interest, since these are theoretically cleanest and, for the \Bsstar\ case, are sensitive to the same operators which could be responsible for the flavour anomalies.\footnote{
  The inclusion of charge conjugate processes is implied throughout this paper.}
These decays have been considered as a potential probe of physics beyond the SM in Refs~\cite{Grinstein:2015aua,Khodjamirian:2015dda}, with further investigations of the impact of particular extensions of the SM considered in Refs.~\cite{Xu:2015eev,Sahoo:2016edx,Banerjee:2017upm,Kumar:2017xgl,Kumbhakar:2018uty}.

There is an extra motivation to search for leptonic $\Dstarz$ decays.
The relevant operators for the $\Bstarz$ and $\Bsstar$ decays are already constrained from measurements of pseudoscalar $B_{(s)} \to \ell^+\ell^-$ and $h\ell^+\ell^-$ transitions (where $h$ is a hadron), and can be used to set limits on ${\cal B}\left(\decay{\BorBsstar}{\ell^+\ell^-}\right)$ that are below the current experimentally achievable sensitivity~\cite{Grinstein:2015aua}.
The interpretation of results from $\Dz \to \ell^+\ell^-$ and $D_{(s)} \to h\ell^+\ell^-$ decays, on the other hand, is challenging due to long-distance effects~\cite{Burdman:2001tf,deBoer:2015boa,Bause:2019vpr,Bharucha:2020eup}.  
Correspondingly, constraints on the relevant operators are weaker, and the possibility of a large enhancement to ${\cal B}\left(\decay{\Dstar}{\ell^+\ell^-}\right)$ from physics beyond the Standard Model cannot be ruled out.
Thus, if sufficient experimental precision can be obtained, searches for $\Dstarz\to\ell^+\ell^-$ decays can provide an important complementary approach to probe the operators involved, and may reveal exciting results.

Since there is no suppression of the coupling of the heavy flavoured vector resonances to dielectron, compared to dimuon, states, it may be attractive experimentally to search for these interactions through production in $\epem$ collisions, rather than in decay processes~\cite{Khodjamirian:2015dda}.
A search for the $e^+e^- \to D^{*}(2007)^0$ process has been carried out by the CMD-3 collaboration, resulting in an upper limit being set, ${\cal B}\left(\Dstarz \to \epem \right) < 1.7 \times 10^{-6}$ at 90\% confidence level~\cite{Shemyakin:2020uye}.
While it is likely that this result can be significantly improved through analyses of larger data samples with better background suppression, the limit is not at such a stringent level to suggest that other approaches to study these processes would be futile.

In particular, the copious production of heavy flavoured hadrons at the LHC makes it worthwhile to consider what sensitivity might be achievable with current and future data samples.
As the dimuon signature allows for effective background suppression in hadron collider experiments, the $\Dstarz$, $\Bstarz$ and $\Bsstar$ decays to the $\mu^+\mu^-$ final state are the most amenable.
There are several possible avenues to investigate these processes using LHC data.
In particular, both ``prompt'' and ``displaced'' production can be considered, where in the former the heavy flavoured state is produced at the primary vertex of the proton-proton collision while in the latter the signal hadron originates from the decay of another weakly decaying particle some distance from the primary vertex.
Specifically, $\Dstarz$ mesons are produced at high rates in $b$ hadron decays, and $\BorBsstar$ mesons can be produced in $\Bc$ meson, and potentially in other doubly-heavy hadron, decays. 
While prompt production has the highest rate, the large numbers of tracks originating from the proton-proton collision vertices lead to large backgrounds that limit the sensitivity of any rare decay search.
In displaced production it is unlikely to find random tracks appearing to come from the vertex position, so long as the vertex resolution is sufficient. 
If in addition the signal is part of an exclusive decay process, additional constraints can be applied to suppress further the background.
Displaced production therefore appears to provide the most promising approach. 

In the main part of this paper, the potential sensitivity of the LHCb experiment to $\Dstarz$, $\Bstarz$ and $\Bsstar$ decays to the $\mu^+\mu^-$ final state, using displaced production in exclusive final states is investigated.   
Possibilities with prompt production, and with inclusive and semi-inclusive search approaches with displaced production, are considered for completeness in Appendices~\ref{prompt} and~\ref{inclusive}, respectively.
In principle the ATLAS and CMS experiments could also make competitive measurements, but until now they have fewer relevant measurements making extrapolations difficult.
Moreover, as their vertexing and particle identification capability is not as good as that of LHCb, it is expected that they will suffer from larger backgrounds.

\section*{\boldmath $\Bm \to \Dstarz \pim \to \mumu \pim$ decays}
\begin{figure}[!tb]
  \centering
  \includegraphics[width=0.7\linewidth]{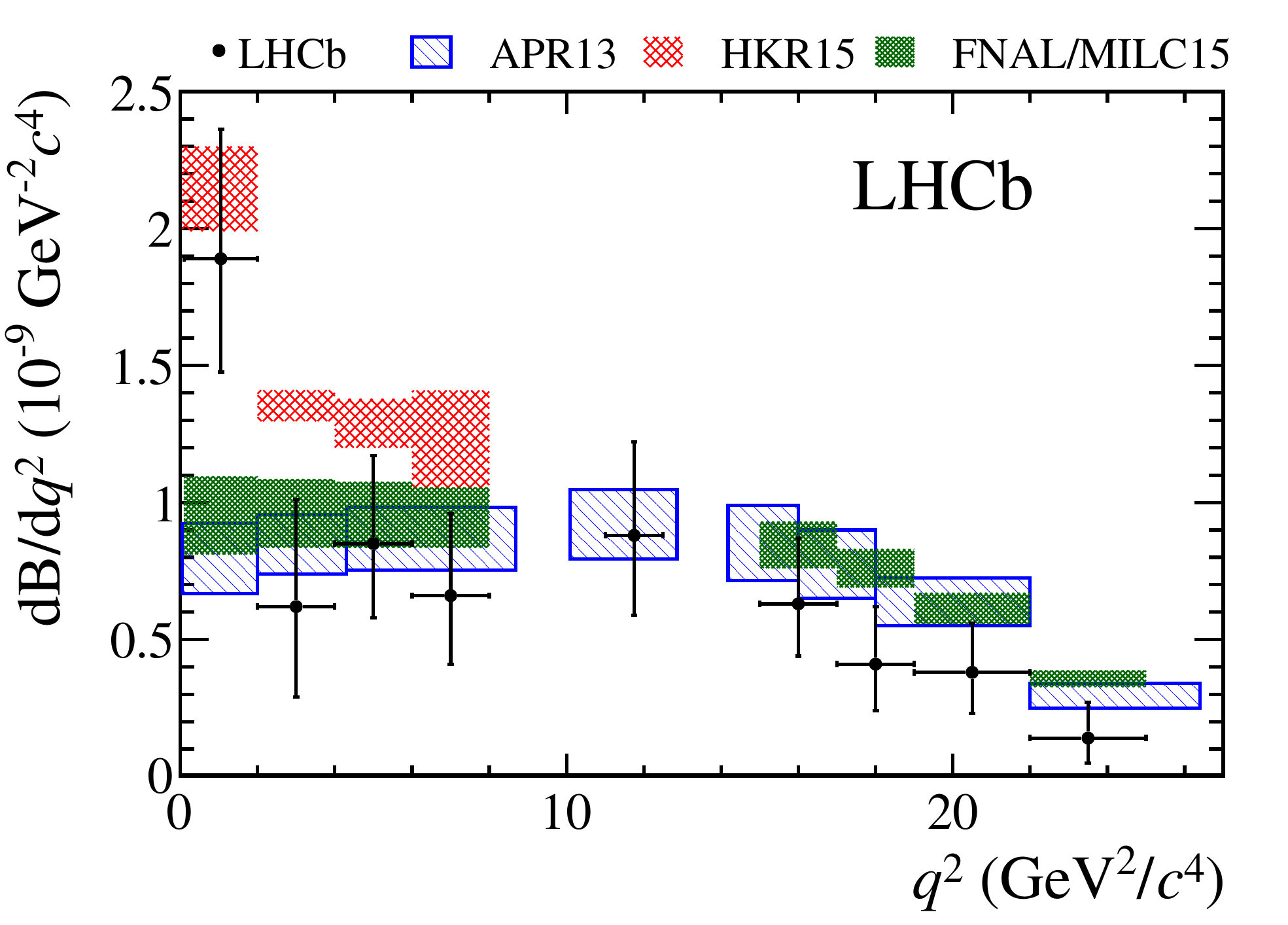}
  \caption{Differential branching fraction of the $\Bm \to \mumu \pim$ decay as a function of $\qsq = m(\mumu)^2$, taken from Ref.~\cite{LHCb-PAPER-2015-035}.
  The hashed regions correspond to theoretical predictions from Refs.~\cite{Ali:2013zfa,Hambrock:2015wka,FermilabLattice:2015cdh}.}
  \label{fig:pimumu}
\end{figure}

The decay chain $\Bm \to \Dstarz \pim \to \mumu \pim$ not only provides an excellent illustrative example, it also allows a limit on the branching fraction for $\Dstarz \to \mumu$ to be set from published data.
In particular, LHCb has studied the $\Bm \to \mumu \pim$ decay using a data sample corresponding to an integrated luminosity of $3.0 \invfb$, collected at centre-of-mass energies of 7 and $8 \tev$~\cite{LHCb-PAPER-2015-035}.
The measured differential branching fraction ${\rm d}{\cal B}/{\rm d}\qsq$, where $\qsq$ is the square of the dimuon invariant mass, is shown in Fig.~\ref{fig:pimumu}.
An upper limit on ${\cal B}\left(\Dstarz \to \mumu \right)$ can be set by assuming conservatively that the $\Bm \to \Dstarz \pim \to \mumu \pim$ decay contributes not more than half of the $\Bm \to \mumu \pim$ signal in the two bins either side of $\qsq = 4 \gevgevcccc$, {\it i.e.}\ a branching fraction of ${\cal B}\left( \Bm \to \left[ \mumu \right]_{\Dstarz} \pim\right) \lsim 1.5 \times 10^{-9}$.
Then, using the world average value of ${\cal B}\left( \Bm \to \Dstarz \pim \right)$~\cite{PDG2020}, one obtains
\begin{equation*}
  {\cal B}\left( \Dstarz \to \mumu \right) = \frac{{\cal B}\left( \Bm \to \left[ \mumu \right]_{\Dstarz} \pim\right)}{{\cal B}\left( \Bm \to \Dstarz \pim\right)} \lsim 3 \times 10^{-7} \, .
\end{equation*}
This limit, while already more stringent than the result on $\Dstarz \to \epem$ from the CMD-3 collaboration~\cite{Shemyakin:2020uye}, could most likely be improved by at least an order of magnitude by a dedicated LHCb analysis.
In particular, the experimental mass resolution, which we expect to be around $5 \mevcc$, is much better than can be obtained from the coarse $\qsq$ binning of Fig.~\ref{fig:pimumu}.
Moreover, LHCb has already collected a significantly larger data sample than was analysed in Ref.~\cite{LHCb-PAPER-2015-035}, and a somewhat higher selection efficiency could be anticipated in a dedicated analysis.
It is therefore of interest to ask what sensitivity might ultimately be achieved by LHCb in such a search.

In LHC experiments, it is advantageous to measure branching fractions relative to those of appropriate normalisation channels.
It is anticipated that the \decay{\Bm}{\jpsi\Km} decay will be used for this purpose and hence the signal yield is converted to a measurement of the branching fraction as
\begin{eqnarray}
    {\cal B}\left(\decay{\Dstarz}{\mumu}\right) & = &
    \frac{N_{\decay{\Bm}{\Dstarz\pim}}}{N_{\decay{\Bm}{\jpsi\Km}}} \,
    \frac{\epsilon_{\decay{\Bm}{\jpsi\Km}}}{\epsilon_{\decay{\Bm}{\Dstarz\pim}}} \,
    \frac{{\cal B}\left(\decay{\Bm}{\jpsi\Km}\right)}{{\cal B}\left(\decay{\Bm}{\Dstarz\pim}\right)} \, 
    {\cal B}\left(\decay{\jpsi}{\mumu}\right) \, , \label{BFnorm} \\
    & = & \alpha_{\decay{\Dstarz}{\mumu}} N_{\decay{\Bm}{\Dstarz\pim}} \, ,
\end{eqnarray}
where $\alpha_{\decay{\Dstarz}{\mumu}}$ is the single-event sensitivity, which corresponds to the branching fraction at which one signal event is expected in the data sample.
In Eq.~\eqref{BFnorm}, $N$ and $\epsilon$ represent the yield and the efficiency for the decay indicated in the subscript, where the reconstruction of the \Dstarz\ or \jpsi\ vector meson in the \mumu\ final state is implied.

\begin{figure}[!tb]
    \centering
    \includegraphics[width=0.7\linewidth]{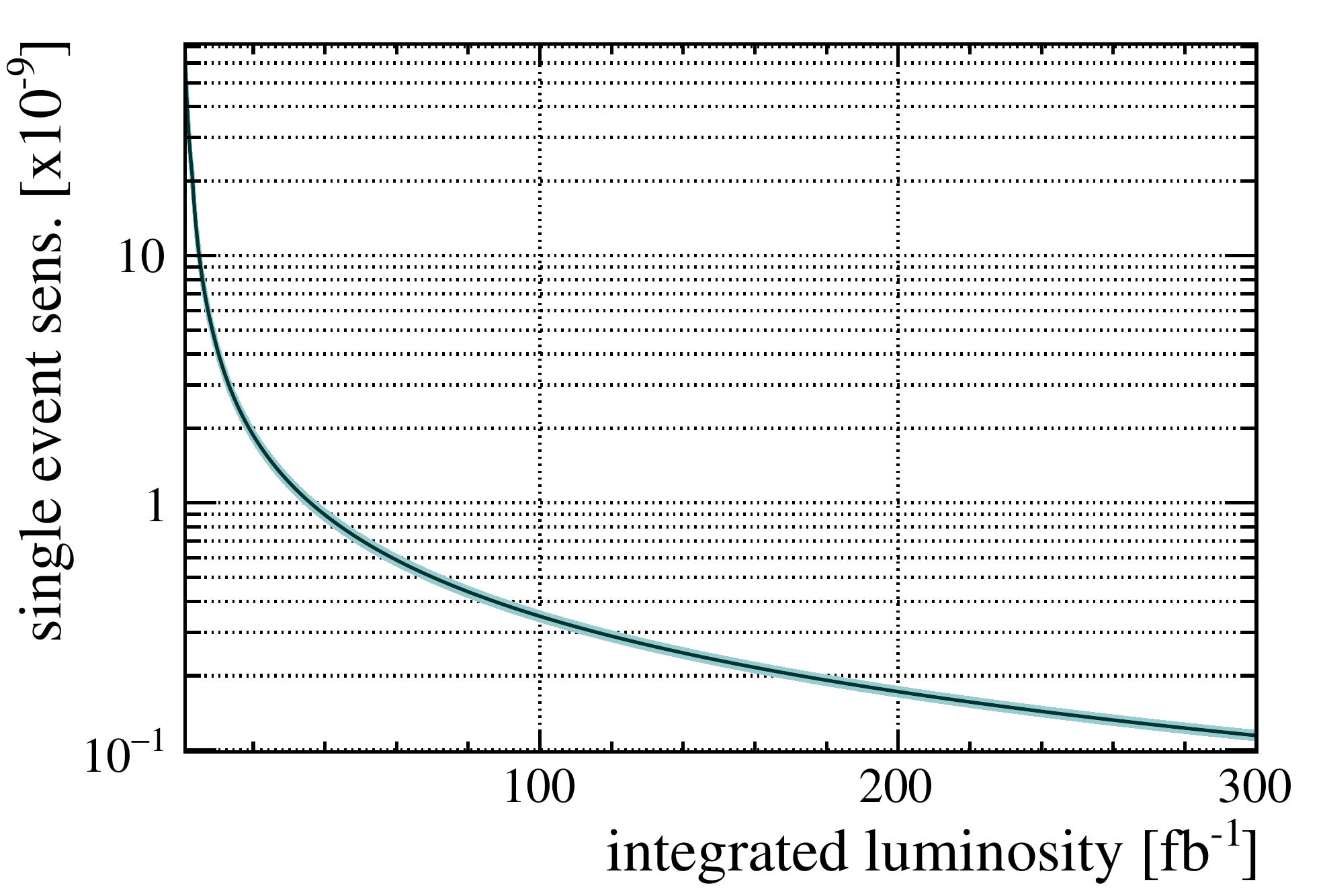}
    \caption{
    Expected single-event sensitivity for \Dstarz produced in \decay{\Bm}{\Dstarz\pim} decays as a function of the integrated luminosity of the LHCb data set.
    }
    \label{fig:ses:Btopimumu}
\end{figure}

The single-event sensitivity is estimated from the yield of \decay{\Bm}{\jpsi\Km} decays, scaled from measurements with existing data~\cite{LHCb-PAPER-2015-035} appropriately to each integrated luminosity value, and the branching fractions $\BF(\decay{\Bm}{\jpsi\Km})$, $\BF(\decay{\jpsi}{\mumu})$ and $\BF(\decay{\Bm}{\Dstarz\pim})$~\cite{PDG2020}. 
It is assumed that $\epsilon_{\decay{\Bm}{\jpsi\Km}}\approx\epsilon_{\decay{\Bm}{\Dstarz\pim}}$, since the final state is the same (apart from $K \leftrightarrow \pi$) and the efficiency varies slowly with $\qsq$.
The single-event sensitivity is shown in Fig.~\ref{fig:ses:Btopimumu} as a function of the data set size. 

The achievable precision depends not only on the single-event sensitivity but also on the uncertainty on the signal yield, which is often limited by the background level.
To investigate how the achievable limit may scale with integrated luminosity, pseudoexperiments are generated under a background-only hypothesis. 
Three background components are considered: 
combinatorial background from random combinations of tracks from two or more decays; background from nonresonant \decay{\Bm}{\mumu\pim} decays; and background from \decay{\Bm}{\Km\mumu} decays, where the \Km meson is mistakenly identified as a \pim. 
The combinatorial background is assumed to be uniformly distributed in the dimuon mass, $m(\mumu)$. 
The backgrounds from \decay{\Bm}{\Km\mumu} and \decay{\Bm}{\mumu\pim} decays are assumed to be uniform in $\qsq = m^{2}(\mumu)$, consistent with both the expected~\cite{Ali:1999mm} and the observed~\cite{LHCb-PAPER-2014-006} shape of the differential \decay{\Bm}{\Km\mumu} decay rate in the \qsq range of interest.
The backgrounds are generated over the interval $2 < \qsq < 6\gev^{2}$, which covers the two bins of the LHCb \decay{\Bm}{\mumu\pim} analysis closest to the \Dstarz mass. 
The level of each of the three backgrounds is taken from Ref.~\cite{LHCb-PAPER-2015-035} and is scaled to the considered integrated luminosity.
Proton-proton collisions during LHC Run 2 (2015--18) were at $13 \tev$ centre-of-mass energy and those in future run periods are expected to be at similar or slightly higher energies.
When extrapolating to future data sample sizes it is assumed that the \bbbar production cross-section scales linearly with centre-of-mass energy~\cite{LHCb-PAPER-2016-031}. 

Limits are set for each pseudoexperiment using the method described in Ref.~\cite{Rolke:2004mj} (as implemented in the \texttt{TRolke} class in \root), taking the uncertainty on the single-event sensitivity into account. 
Two mass regions are defined, a signal region $\pm 10\mevcc$ around the known \Dstarz mass,  $m_{\Dstarz}$, comprising a mixture of signal and background candidates and two sideband regions $-35 < m(\mumu) - m_{\Dstarz} < -15\mevcc$ and $15 < m(\mumu) - m_{\Dstarz} < 35\mevcc$, comprising only background candidates. 
The sideband regions are used to estimate the background in the signal region. 
The width of the signal region is taken as $\pm 2$ times the expected $m(\mup\mun)$ resolution of $\sim 5\mevcc$~\cite{LHCb-PAPER-2016-045}, which is minimised by applying a kinematic fit to the \decay{\Bm}{\mumu\pim} process in which the \Bm\ mass is constrained to its known value. 

\begin{figure}[!tb]
    \centering
    \includegraphics[width=0.7\linewidth]{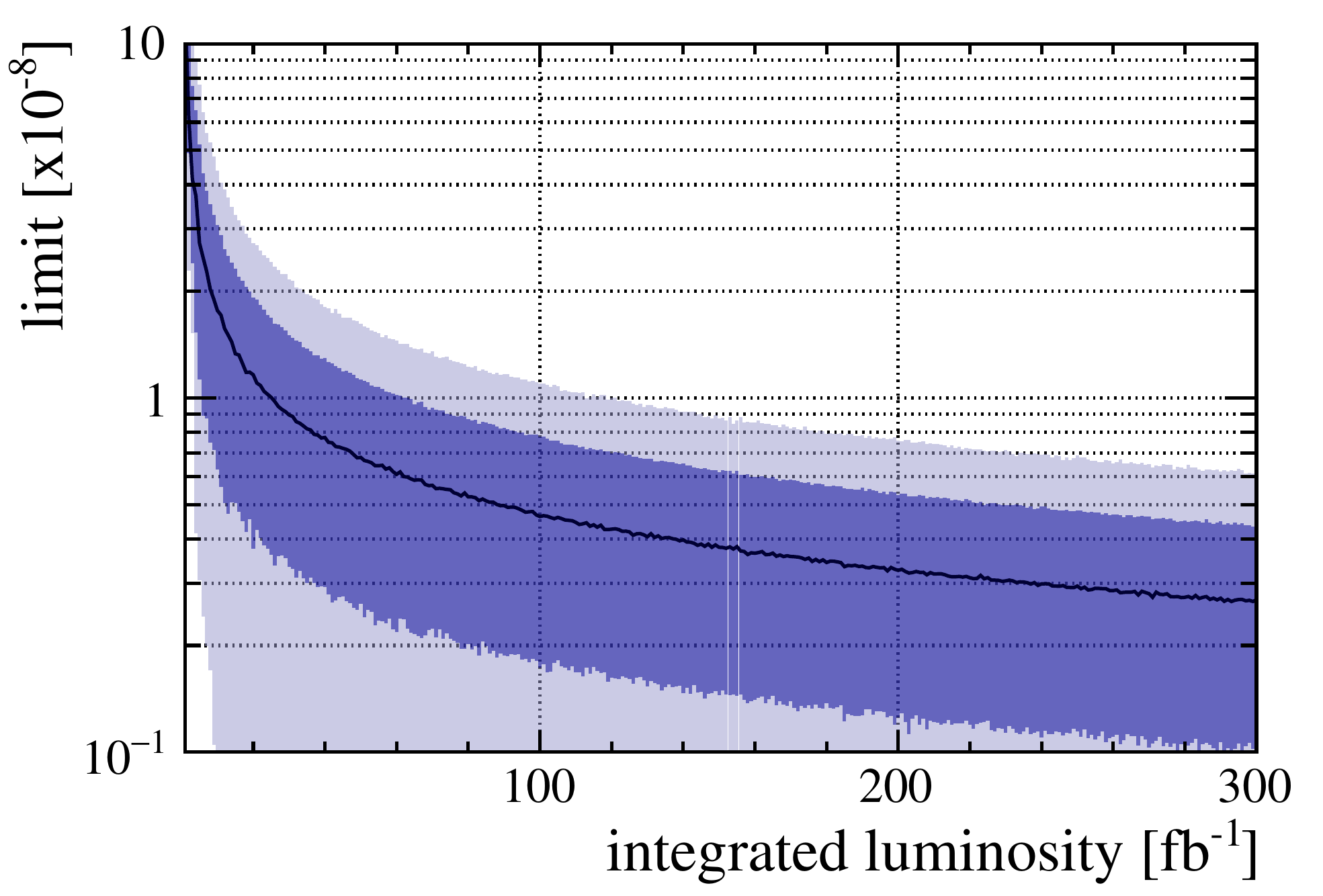}
    \caption{
        Expected upper limit on $\BF(\decay{\Dstarz}{\mumu})$ obtained using reconstructed \decay{\Bm}{\mumu\pim} decays as a function of the integrated luminosity of the LHCb data set. 
        The curve represents the median value from an ensemble of pseudoexperiments and the shaded regions the one and two sigma intervals.}
    \label{fig:limit:Btopimumu}
\end{figure}

The results of this pseudoexperiment-based study are shown in Fig.~\ref{fig:limit:Btopimumu}. 
An upper limit at the level of $10^{-8}$ appears to be possible with the current LHCb data set. 
This can be further improved to ${\cal O}(10^{-9})$ with the total sample of $300 \invfb$ anticipated with future LHCb upgrades~\cite{LHCb-PII-Physics,LHCb-PII-EoI}. 
The rate of reduction of the expected limit as the sample size increases slows markedly at around $20 \invfb$ as background start to become limiting.
While combinatorial background and misidentified \decay{\Bm}{\Km\mumu} decays can be further reduced with tighter selection requirements, the contribution from nonresonant \decay{\Bm}{\mumu\pim} decays is irreducible in this approach.

The studies resulting in Fig.~\ref{fig:limit:Btopimumu} neglect interference effects between the $\decay{\Bm}{\left(\decay{\Dstarz}{\mumu}\right)\pim}$ signal and nonresonant $\decay{\Bm}{\pim\mumu}$ decays, which is reasonable given the narrow width of the \Dstarz\ meson.
Nonetheless, since the \Dstarz\ quantum numbers are the same as part of the nonresonant contribution, a net interference effect is expected, with size depending on the relative phase between the two interfering amplitudes.   
Even though the \Dstarz\ width is below the experimental resolution, this effect could in principle be used, once the sample size is sufficiently large, to obtain a better limit than indicated in Fig.~\ref{fig:limit:Btopimumu}.
Indeed, the LHCb collaboration has already demonstrated the possibility to measure a similar interference effect between $\decay{\Bm}{\left(\decay{\jpsi}{\mumu}\right)\Km}$ and nonresonant $\decay{\Bm}{\Km\mumu}$ decays~\cite{LHCb-PAPER-2016-045}.
The results of this analysis allow a limit on the $\decay{\Bm}{\Dstarz\Km}$ contribution to $\decay{\Bm}{\Km\mumu}$ decays, and hence a limit on ${\cal B}\left(\decay{\Dstarz}{\mumu}\right)$ can be derived.
Assuming the yield of $\decay{\Bm}{\left(\decay{\Dstarz}{\mumu}\right)\Km}$ decays is not more than $2\times \sqrt{N}$, where $N \sim 50$ is the yield in the $m(\mumu)$ bin around $m_{\Dstarz}$, taken from Fig.~3 of Ref.~\cite{LHCb-PAPER-2016-045}, and normalising to the $\decay{\Bm}{\left(\decay{\jpsi}{\mumu}\right)\Km}$ contribution which has a yield of $\sim 900\,000$ (around 90\% of the total $\decay{\Bm}{\Km\mumu}$ yield), and with an efficiency ratio of 0.85 (from Fig.~2 of Ref.~\cite{LHCb-PAPER-2016-045}), then with known branching fractions~\cite{PDG2020,LHCb-PAPER-2020-036} one obtains
\begin{eqnarray*}
    {\cal B}\left(\decay{\Dstarz}{\mumu}\right) & = &
    \frac{N_{\decay{\Bm}{\Dstarz\Km}}}{N_{\decay{\Bm}{\jpsi\Km}}} \,
    \frac{\epsilon_{\decay{\Bm}{\jpsi\Km}}}{\epsilon_{\decay{\Bm}{\Dstarz\Km}}} \,
    \frac{{\cal B}\left(\decay{\Bm}{\jpsi\Km}\right)}{{\cal B}\left(\decay{\Bm}{\Dstarz\Km}\right)} \, 
    {\cal B}\left(\decay{\jpsi}{\mumu}\right) \, , \\
    & \lsim & 3 \times 10^{-6} \, .
\end{eqnarray*}
As expected, this mode is not as sensitive as $\decay{\Bm}{\left(\decay{\Dstarz}{\mumu}\right)\pim}$ due to the smaller \Bm\ branching fraction and the larger background from nonresonant $\decay{\Bm}{\Km\mumu}$ decays.

\section*{\boldmath $\Bcp \to \BorBsstar \pip \to \mumu \pip$ decays}
A similar strategy can in principle also be employed to search for $\Bcp \to \BorBsstar \pip \to \mumu\pip$ decays, where the $\Bcp \to \jpsi \pip$ decay can be used for normalisation.
In this case, however, some of the necessary ingredients to convert the signal yield to the $\decay{\BorBs}{\mumu}$ branching fraction are not currently available.
Specifically, the equivalent expression to Eq.~\eqref{BFnorm} is 
\begin{eqnarray}
    {\cal B}\left(\decay{\BorBsstar}{\mumu}\right) & = &
    \frac{N_{\decay{\Bcp}{\BorBsstar\pip}}}{N_{\decay{\Bcp}{\jpsi\pip}}} \,
    \frac{\epsilon_{\decay{\Bcp}{\jpsi\pip}}}{\epsilon_{\decay{\Bcp}{\BorBsstar\pip}}} \,
    \frac{{\cal B}\left(\decay{\Bcp}{\jpsi\pip}\right)}{{\cal B}\left(\decay{\Bcp}{\BorBsstar\pip}\right)} \, 
    {\cal B}\left(\decay{\jpsi}{\mumu}\right) \, , \label{BFnorm2} \\
    & = & \alpha_{\decay{\BorBsstar}{\mumu}} N_{\decay{\Bcp}{\BorBsstar\pip}} \, ,
\end{eqnarray}
and the branching fractions for $\Bcp \to \Bsstar \pip$ and $\Bcp \to \Bstarz \pip$ decays are at present unmeasured. 
It seems possible, however, that at least the former decay could be observed with existing data.
A previous LHCb analysis has observed $\Bcp \to \Bs \pip$ with $3\invfb$ of data~\cite{LHCb-PAPER-2013-044} and this channel has also been used with the full current data sample of $9\invfb$ to measure the $\Bc$ mass~\cite{LHCb-PAPER-2020-003}.
The signal for $\Bcp \to \Bsstar \pip$ would be expected in the same spectrum as a satellite peak shifted below the $\Bcp$ mass by approximately $m_{\Bsstar} - m_{\Bs}$ due to the soft photon that is not included in the reconstructed candidate.
Similar ``partial reconstruction'' techniques have been used by LHCb to observe $\Bcp \to \jpsi \Dssp$~\cite{LHCb-PAPER-2013-010} and $\Bm \to \Dstarz \pim$~\cite{LHCb-PAPER-2017-021} decays, where the soft neutral particles from the $\Dssp$ and $\Dstarz$ decays are not included in the reconstructed candidate.
Assuming that ${\cal B}(\Bcp \to \Bsstar \pip)$ is not much smaller than ${\cal B}(\Bcp \to \Bs \pip)$, as is expected theoretically~\cite{Ivanov:2006ni,Sun:2008wa,Likhoded:2010jr,Naimuddin:2012dy,Shi:2016gqt},
it should be possible to observe partially reconstructed $\Bcp \to \Bsstar \pip$ decays, and to measure the corresponding branching fraction, in the existing LHCb data sample.
The $\Bcp \to \Bz \pip$ and $\Bcp \to \Bstarz \pip$ decays could be searched for with a similar technique although, since these transitions are Cabibbo-suppressed relative to $\Bcp \to \BsorBsstar \pip$, larger data samples may be required to observe them.
A further, albeit, minor complication is that the $m_{\Bstarz} - m_{\Bz}$ mass difference is as-yet unmeasured, although it can be predicted rather reliably from the $m_{\Bstarp} - m_{\Bp}$ mass difference invoking isospin symmetry. 
(This can also be interpreted as a further motivation for the analysis since a first measurement of $m_{\Bstarz}$ may be possible.)

Another apparent problem is that even for $\Bcp$ decay modes which have been observed, there are currently no measurements of absolute branching fractions.
Rather, only the product of the branching fraction with a ratio of production fractions is known.
However, the $\Bc$ meson production rate cancels out in Eq.~\eqref{BFnorm2}, allowing some simplification.
With ${\cal B}\left(\decay{\Bcp}{\jpsi\pip}\right)$, which appears in the numerator, measured relative to \decay{\Bp}{\jpsi\Kp} and ${\cal B}\left(\decay{\Bcp}{\BorBsstar\pip}\right)$, which appears in the denominator, measured relative to \decay{\Bs}{\jpsi\phi}, it is only necessary to know the relative production rate of $\Bp$ and $\Bs$ mesons, which has been measured precisely~\cite{LHCb-PAPER-2020-046}.
Taking the yield of $25.2 \times 10^3$ $\decay{\Bcp}{\jpsi\pip}$ decays in $9 \invfb$ of LHCb data from Ref.~\cite{LHCb-PAPER-2020-003} and other inputs from Refs.~\cite{PDG2020, LHCb-PAPER-2013-044,LHCb-PAPER-2014-050,LHCb-PAPER-2020-046}, and assuming ${\cal B}\left( \decay{\Bcp}{\Bsstar\pip}\right) = \frac{1}{2}{\cal B}\left( \decay{\Bcp}{\Bs\pip}\right)$ and that the ratio of efficiencies in Eq.~\eqref{BFnorm2} is unity, a single-event sensitivity with $9 \invfb$ of $\alpha_{\decay{\Bsstar}{\mumu}} \approx 1.4 \times 10^{-8}$ is obtained.  
Further assuming the \decay{\Bcp}{\Bstarz\pip} decay has a Cabibbo-suppression factor of $\left|V_{cd}/V_{cs}\right|^2 \approx 5\%$ relative to \decay{\Bcp}{\Bsstar\pip}, the corresponding single-event sensitivity is found to be $\alpha_{\decay{\Bstarz}{\mumu}} \approx 2.7 \times 10^{-7}$.
Scaling to a data sample of $300 \invfb$, the achievable single-event-sensitivities could reach $\approx 3.4 \times 10^{-10}$ and $6.8 \times 10^{-9}$ for \decay{\Bsstar}{\mumu} and \decay{\Bstarz}{\mumu} decays, respectively.

Interpretation of these single-event sensitivities must be made with care, however, since they assume a selection efficiency comparable to that for $\decay{\Bcp}{\jpsi\pip}$ decays in Ref.~\cite{LHCb-PAPER-2020-003}.
In practice the selection requirements will be optimised to account for the level of background in the signal region for the $\Bcp \to \BorBsstar \pip \to \mumu \pip$ search.
Since there is no published search for $\Bcp \to \mumu \pip$ decays outside the mass regions where the dimuon pair originates from a $\jpsi$ or $\psi(2S)$ charmonium state, it is hard to judge what the optimal requirements are likely to be.
Moreover a contribution from nonresonant $\Bcp \to \mumu \pip$ decays, 
which can occur in the SM through an annihilation diagram, may provide a limiting background if it is not negligible around $m(\mumu) \sim m_{\BorBsstar}$.
Nevertheless, it appears possible that interesting sensitivity to \decay{\Bsstar}{\mumu} and \decay{\Bstarz}{\mumu} decays may be achievable.

\section*{Summary}
In summary, the \decay{\Bm}{\mumu\pim} and \decay{\Bcp}{\mumu\pip} decays provide interesting possibilities to search for the leptonic weak decays of \Dstarz, \Bstarz\ and \Bsstar\ vector mesons. 
Published data allow a world-leading limit of ${\cal B}\left(\decay{\Dstarz}{\mumu}\right) \lsim 3 \times 10^{-7}$ to be obtained, and this can be significantly improved upon with a dedicated analysis of existing data.
Sensitivity at the level of ${\cal O}(10^{-9})$ is expected to be possible with the data sample to be collected by the end of HL-LHC operation with upgrades of the LHCb experiment.
Good sensitivity to \decay{\BorBsstar}{\mumu} decays also appears achievable, although further experimental investigations will be needed before a firm conclusion can be reached on this point.
In particular, measurements of ${\cal B}\left(\decay{\Bcp}{\BorBsstar\pip}\right)$ and studies of nonresonant \decay{\Bcp}{\mumu\pip} decays are needed.

\section*{Acknowledgements}

The authors wish to thank their colleagues on the LHCb experiment for the fruitful and enjoyable collaboration that inspired this study.
In particular, they would like to thank Niels Tuning and Mike Williams for helpful comments.
Additionally, the authors acknowledge stimulating discussions with Benjam\'\i{}n Grinstein and Alexey Petrov.
This work is supported by the Royal Society and the Science and Technology Facilities Council (UK) and the Monash Warwick Alliance.

\appendix
\section{Sensitivity of searches exploiting prompt production}
\label{prompt}

The predominant fraction of heavy-flavoured mesons are produced ``promptly'' in LHC collisions, {\it i.e.}\ they originate directly from the $pp$ interaction vertex (the so-called primary vertex).
Thus one might wonder whether the large production rate could overcome the sizeable background from random combinations of tracks produced at the primary vertex, and allow competitive results on \Dstarz, \Bstarz\ and \Bsstar\ dimuon decays to be obtained.

\begin{figure}[tb]
  \centering
  \includegraphics[width=0.98\linewidth]{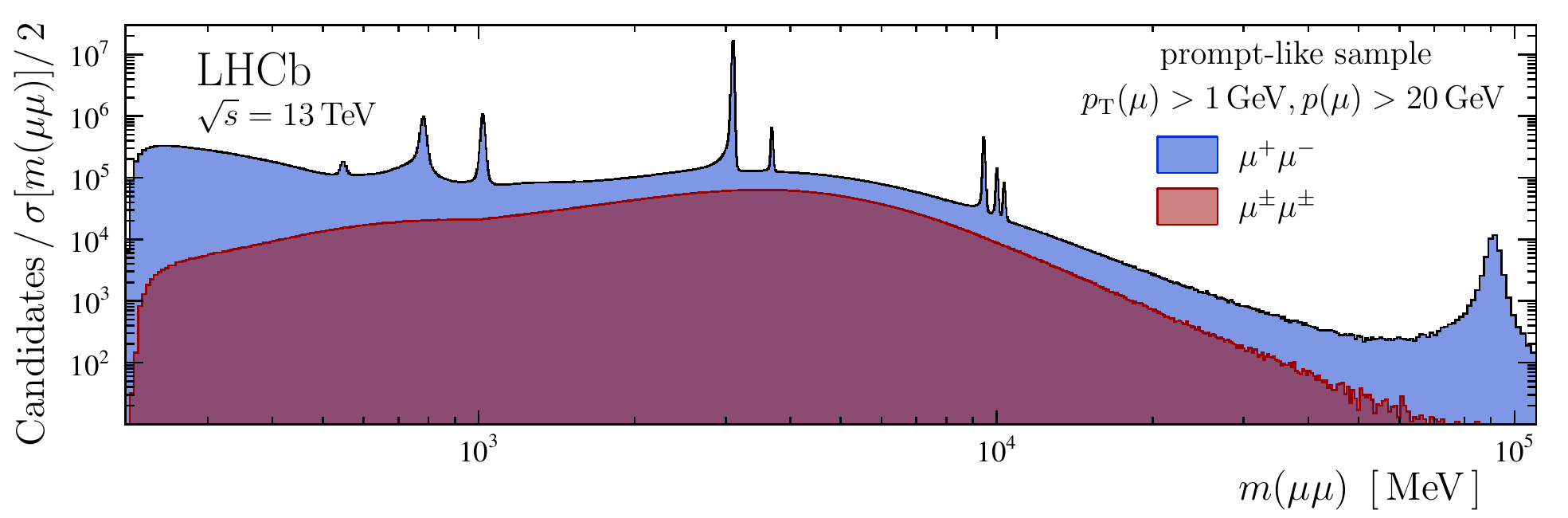}
  \caption{Spectrum of prompt dimuon pairs recorded by LHCb, with background contributions indicated, from Ref.~\cite{LHCb-PAPER-2017-038}.}
  \label{fig:darkphoton}
\end{figure}

Fortuitously, the LHCb collaboration has published, as part of a search for dark photons decaying to $\mumu$~\cite{LHCb-PAPER-2017-038}, the prompt dimuon spectrum obtained from a data sample corresponding to $1.6 \invfb$ of proton-proton collisions recorded at a centre-of-mass energy of $13 \tev$, as shown in Fig.~\ref{fig:darkphoton}.\footnote{
    The spectrum used to obtain limits in Ref.~\cite{LHCb-PAPER-2017-038} includes additional criteria to select preferentially candidates from Drell-Yan production processes, which are not appropriate here.
    This is also the case in Ref.~\cite{LHCb-PAPER-2019-031}, which is based on a larger data sample.
    A spectrum of prompt dimuons without any additional requirements, appropriate for the search proposed here, is available in Ref.~\cite{LHCb-PAPER-2020-013}; however peaks from the $\jpsi$ and other resonances have been vetoed so that no normalisation sample is available in the same spectrum.
}
This can be used to estimate the sensitivity to the dimuon decays for each of the \Dstarz, \Bstarz\ and \Bsstar\ meson.
As an example, for the \Dstarz\ case,
\begin{eqnarray}
    {\cal B}\left(\decay{\Dstarz}{\mumu}\right) & = &
    \frac{N_{\decay{\Dstarz}{\mumu}}}{N_{\decay{\jpsi}{\mumu}}} \,
    \frac{\epsilon_{\decay{\jpsi}{\mumu}}}{\epsilon_{\decay{\Dstarz}{\mumu}}} \,
    \frac{\sigma_{\jpsi}}{\sigma_{\Dstarz}}\, 
    {\cal B}\left(\decay{\jpsi}{\mumu}\right) \, , \label{BFnorm-prompt} \\
    & = & \alpha^{\rm prompt}_{\decay{\Dstarz}{\mumu}} N_{\decay{\Dstarz}{\mumu}} \, ,
\end{eqnarray}
where $N$ and $\epsilon$ represent the yields for the decays given in the subscripts. 
Since analysis of data from the LHCb experiment is being considered, $\sigma_{\jpsi}$ and $\sigma_{\Dstarz}$ refer to the inclusive prompt $\proton\proton\to \jpsi$ and $\proton\proton\to\Dstarz$ production cross-sections in the LHCb acceptance.  

The \jpsi\ production cross-section at $\sqrt{s} = 13 \tev$ has been measured to be $\sigma_{\jpsi} = 15.03 \pm 0.03 \pm 0.94 \mub$~\cite{LHCb-PAPER-2015-037}. 
The cross-section for the \Dstarz\ meson is not measured, but the approximation $\sigma_{\Dstarz} \sim \sigma_{\Dstarp}$ can be made, so that $\sigma_{\Dstarp} = 784 \pm 4 \pm 87 \mub$ at $13 \tev$~\cite{LHCb-PAPER-2015-041} can be used.
Hence, estimating $N_{\decay{\jpsi}{\mumu}} \sim 2 \times 10^7$ from Fig.~\ref{fig:darkphoton}, and assuming that the ratio of efficiencies is approximately unity, one obtains a single-event sensitivity with $1.6 \invfb$ of $13 \tev$ data of  $\alpha^{\rm prompt}_{\decay{\Dstarz}{\mumu}} \sim 6 \times 10^{-11}$.

The single-event sensitivities for dimuon decays of \Bstarz\ and \Bsstar\ mesons can be estimated by replacing $\sigma_{\Dstarz}$ with the corresponding \BorBsstar\ production cross-section.  
Since these are not measured, rough estimates need to be made: $\sigma_{\Bstarz}$ is taken to be one-tenth of $\sigma_{\Dstarz}$, with $\sigma_{\Bsstar}$ suppressed by a further factor of four, consistent with the relative production rates of beauty and charm pseudoscalar mesons~\cite{LHCb-PAPER-2015-041,LHCb-PAPER-2017-037,LHCb-PAPER-2020-046}.
This gives $\alpha^{\rm prompt}_{\decay{\Bstarz}{\mumu}} \sim 6 \times 10^{-10}$ and $\alpha^{\rm prompt}_{\decay{\Bsstar}{\mumu}} \sim 2 \times 10^{-9}$.

As is clear from Fig.~\ref{fig:darkphoton}, searches for rare processes in prompt production will suffer from large backgrounds.  
The number of dimuon candidates per bin in the region around $m_{\Dstarz}$ is just over $10^5$, and since the bin width is one half of the mass resolution, this corresponds to a background yield of around $5 \times 10^5$ in a signal window of $\pm 2$ times the mass resolution around the signal peak position.  
Hence the uncertainty on $N_{\decay{\Dstarz}{\mumu}}$ arising from a fit to this distribution is expected to be around $700$.
Similar uncertainties are estimated also for $N_{\decay{\Bstarz}{\mumu}}$ and $N_{\decay{\Bsstar}{\mumu}}$.
Assuming that the upper limit corresponds to twice this uncertainty then gives 
\begin{eqnarray*}
{\cal B}\left(\decay{\Dstarz}{\mumu}\right) & \lsim & 10^{-7} \,, \\
{\cal B}\left(\decay{\Bstarz}{\mumu}\right) & \lsim & 10^{-6} \,, \\
{\cal B}\left(\decay{\Bsstar}{\mumu}\right) & \lsim & 3 \times 10^{-6} \,,
\end{eqnarray*}
where the uncertainty on the single-event sensitivity has not been included.  

These values are rough estimates of the sensitivity that could be obtained from the $1.6 \invfb$ data sample of Ref.~\cite{LHCb-PAPER-2017-038}.
Real measurements with this approach would require the relevant production cross-sections to be determined, rather than approximated as done above.
Such measurements require the reconstruction of the soft neutral particles emitted in the $\Dstarz \to \Dz\piz$ or $\Dz\gamma$, $\Bstarz \to \Bz\gamma$ and $\Bsstar \to \Bs\gamma$ decays, for which the LHCb detector is not optimised.
These challenges are not insurmountable, as demonstrated by LHCb measurements of $\chi_{c}$ and $\chi_{b}$ production with photons reconstructed either in the calorimeter or through $\gamma \to \epem$ conversions in detector material~\cite{LHCb-PAPER-2011-030,LHCb-PAPER-2012-015,LHCb-PAPER-2014-031}.
Nonetheless, given that the limits are background dominated and will therefore improve only slowly with larger data samples, this approach looks less attractive compared to production through \B-meson decays.

\section{Sensitivity of (semi-)inclusive searches exploiting displaced production}
\label{inclusive}

A possible strategy to avoid the large background in prompt searches could be to make an inclusive search for dimuon \Dstarz, \Bstarz\ and \Bsstar\ decays originating from a displaced vertex (\ie\ from $b$-hadron decays).
Considering for example the $\decay{\Dstarz}{\mumu}$ case, the number of signal decays in the sample will be given by 
\begin{equation}
    N_{\decay{\Dstarz}{\mumu}} = 
    {\cal L} \, {\cal B}\left(\decay{\Dstarz}{\mumu}\right)
    \sum_i \sigma_{B_i} \, {\cal B}\left(\decay{B_i}{\Dstarz {\cal X}}\right) \, \epsilon_{\decay{B_i}{\Dstarz {\cal X}}} \, ,
\end{equation}
where ${\cal L}$ is the integrated luminosity of the sample, $\sigma_{B_i}$ is the production cross-section for $b$-hadron species $B_i$ that has an inclusive branching fraction to decay to a final state containing a \Dstarz\ meson of ${\cal B}\left(\decay{B_i}{\Dstarz {\cal X}}\right)$, where ${\cal X}$ indicates any set of additional particles.\footnote{
  Most generally one should consider all weakly decaying hadrons with non-zero branching fractions for decays including \Dstarz\ mesons in the final state.
  This would include double charm baryons, for example.
  However, the additional contributions to the total inclusive rate are negligible.
}
The corresponding efficiency is $\epsilon_{\decay{B_i}{\Dstarz {\cal X}}}$, which will depend on the requirements imposed to select displaced vertices and on the lifetime and kinematics of the $B_i$ hadron.
Considering every possible $B_i$ species would be challenging, but for the \Dstarz\ case production through \Bp\ and \Bz\ decays can be assumed to dominate, and knowledge exists of their production cross-sections~\cite{LHCb-PAPER-2017-037} and inclusive branching fractions~\cite{CLEO:1997kfu}.
As usual for measurements in hadron collider experiments, it is experimentally convenient to use a normalisation channel in order to cancel the luminosity and avoid dependence on the absolute efficiencies.  
The $\jpsi \to \mumu$ channel is again well-suited for this purpose. Contributions from $\Bs$ and $\Lb$ decays are not negligible in this case but measurements exist of all needed quantities~\cite{PDG2020}. 
This approach however is not currently viable for dimuon \Bstarz\ and \Bsstar\ decay searches, due to lack of knowledge of the relevant cross-sections and branching fractions.

\begin{figure}[tb]
  \centering
  \includegraphics[width=0.98\linewidth]{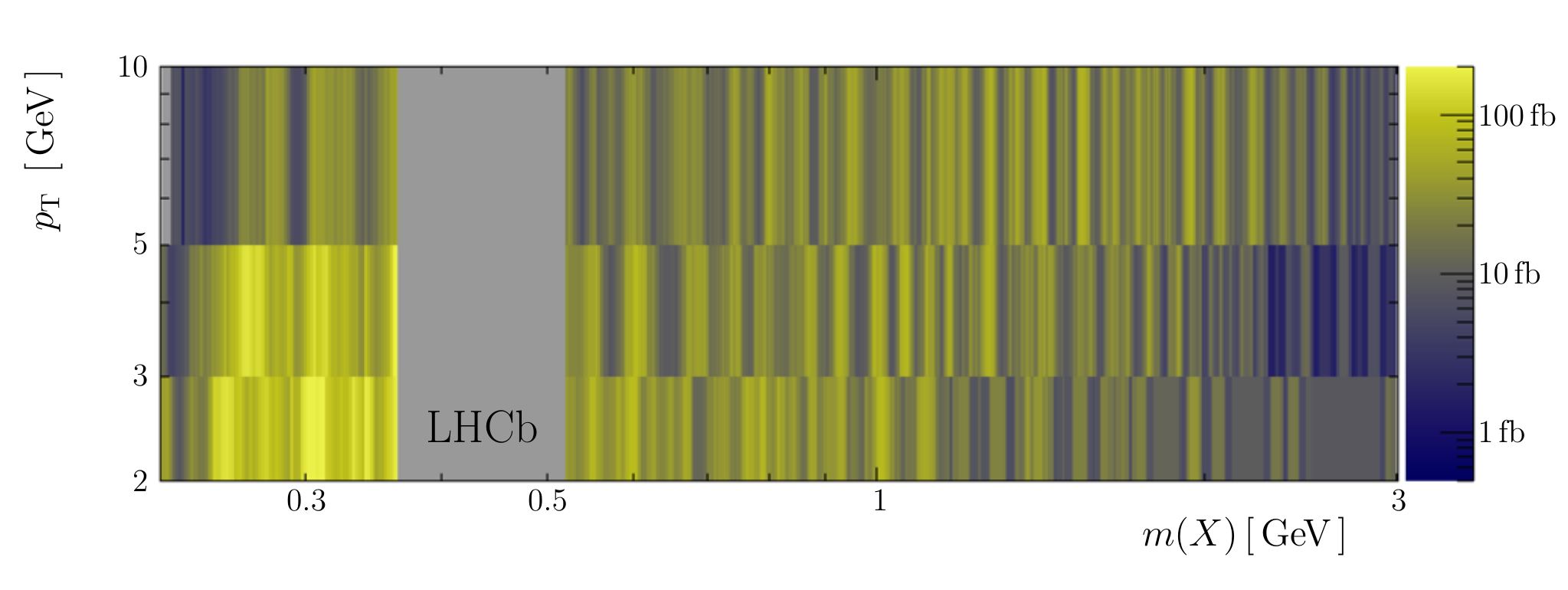}
  \caption{Upper limits at 90\% confidence level on the production cross-section times branching fraction $\sigma(X) \, {\cal B}\left(X\to \mumu\right)$ obtained from an inclusive search for displaced $X\to\mumu$ decays, from Ref.~\cite{LHCb-PAPER-2020-013}.
  Limits are shown as a function of the mass of the hypothesised $X$ particle and its transverse momentum.}
  \label{fig:Xtomumu}
\end{figure}

The LHCb collaboration has published the results of a search for inclusive displaced $X \to \mumu$ decays, without any requirement that the $X$ particle is produced at the primary $pp$ collision vertex~\cite{LHCb-PAPER-2020-013}.
This search was motivated by the possible existence of beyond Standard Model particles (generically denoted $X$) and includes selection requirements to suppress background from dimuon pairs produced in $b$-hadron decay chains.
Moreover, only the region below the $\jpsi$ mass is included in the search so the normalisation channel is not available.  
Therefore, it is not useful to attempt to use the results of this search, shown in Fig.~\ref{fig:Xtomumu}, to obtain information about \decay{\Dstarz}{\mumu} decays; nonetheless, Ref.~\cite{LHCb-PAPER-2020-013} demonstrates that a dedicated analysis using this approach could be carried out.  
However, compared to the exclusive approach emphasised in this paper, the mass resolution will be worse and higher irreducible backgrounds are expected as all $b$-hadron decays producing a nonresonant $\mumu$ pair plus one or more extra particles will contribute.
As such, this approach appears unlikely to be competitive with large data samples.

Background from nonresonant $\mumu$ pairs produced in $b$-hadron decays is irreducible in a fully inclusive search.
However, a semi-inclusive approach could be used to remove it, in which an additional muon (or electron) is required to originate from the same vertex.
In this scenario, the source of \Dstarz\ mesons is through semileptonic $b$-hadron (dominated by \Bp\ and \Bz) decays.
Also in this case there is a relevant prior LHCb publication, in the form of a search for \decay{\Bp}{\mumu\mup\nu} decays~\cite{LHCb-PAPER-2018-037}.
This analysis, however, includes a requirement that one $\mumu$ pair has mass below $980 \mevcc$, and therefore cannot be used directly to obtain a limit on $\decay{\Bp}{\left( \decay{\Dstarzb}{\mumu} \right)\mup\nu}$.
Nonetheless, it appears likely that a dedicated analysis, focusing on the $m(\mumu) \sim m_{\Dstarz}$ mass range, could achieve a limit on ${\cal B}\left( \Dstarz \to \mumu \right)$ of ${\cal O}\left( 10^{-7} \right)$.
The possibility to relax selection requirements to allow additional particles, beyond the neutrino, in the decay could also be exploited if the corresponding improvement in single-event sensitivity is sufficient to overcome the increase in background.
Thus, if background levels can be kept very low, this approach may in the long-term be competitive with, or even better than, the method with exclusive reconstruction of hadronic decays.
Unfortunately, with the currently available information the achievable sensitivity cannot be reliably quantified.
Similarly, the use of \decay{B_i}{\BorBsstar \mu {\cal X}} production, with the most important contribution expected to be from semileptonic \Bc\ decays, to search for \decay{\BorBsstar}{\mumu} transitions appears an interesting possibility, but in the absence of relevant experimental results the achievable sensitivity cannot be estimated.

\addcontentsline{toc}{section}{References}
\setboolean{inbibliography}{true}
\bibliographystyle{LHCb}
\bibliography{references,main,LHCb-PAPER,LHCb-CONF,LHCb-DP,LHCb-TDR}

\end{document}